\documentclass[12pt]{article}

\usepackage{amsfonts}
\usepackage{epsfig}

\begin{document}

\title{ {Probabilistic Embedding Of Discrete Sets As Continuous Metric Spaces}}

\vspace{1cm}

\author{ {Ph. Blanchard \& D. Volchenkov}\footnote{ { E-Mail: volchenk@physik.uni-bielefeld.de} }
\vspace{0.5cm}\\
{ { \small Bielefeld University, Postfach 100131,D-33501, Bielefeld, Germany}}}

\date{ {\today}}
\maketitle

\begin{abstract}
 { Any symmetric affinity function $w: V\times V \to
\mathbb{R}_+$ defined on a discrete set $V$ induces  Euclidean
space structure on $V$. In particular, an undirected graph
specified by an affinity (or adjacency ) matrix can be
considered as a metric topological space. We have calculated the
visual representations of the probabilistic locus for a chain, a
polyhedron, and a finite 2-dimensional lattice. }
\end{abstract}

\vspace{0.5cm}


\section{ {Introduction}}
\label{sec:Introduction}
\noindent

 {
Flatness is
the essential property of Euclidean space.
In particular, there is no distinguished
point that serves as an origin,
because it can be translated anywhere.
This affine-geometric property of Euclidean space
is generalized by the affine group
of all invertible affine transformations from the space into itself
consisting of a linear transformation
described by a matrix $\mathcal{A}$
 followed by a translation,
\[
\bf{y}\,=\, \mathcal{A}\bf{x}+ \bf{a}.
\]
It is known that the affine geometry keeps the concepts of
straight lines and parallel lines but not those of distance
between points or value of angles, \cite{Busemann}.}

 { From a probabilistic point of view,
the fact
that there is
no canonical choice of
 the origin
can be interpreted as if
it can be at any point
in the space with a uniform probability.
The interesting question arises
when the probability $\pi_x>0$
 to find the origin
at the particular point
$x\in V$ of some discrete set
 $V$ of $N$ points
is not uniform,
\begin{equation}
\label{normalization}
\sum_{x\in V}\pi_x\,=\,1,
\end{equation}
-- What does such a space look like?}

 { One way to describe the disparity
 between points in the
set $V$ is by introducing
 an affinity function
  $w: V\times V \to \mathbb{R}_+ \cup \{0\}$
encoding their pairwise relations or neighborhoods - two points
$x\in V$ and $y\in V$ are neighbors, $x\sim y$, iff $w_{xy}>0$,
but $w_{xy}=0$ otherwise. Then, the probability
 distribution $\pi$ is given by
\begin{equation}
\label{stationary_pi}
\pi_x\,=\,\frac{\,\,\sum_{y\in V}w_{xy}\,\,}{\,\,\sum_{y,z\in V}w_{yz}\,\,}.
\end{equation}
The affinity function $w_{xy}\geq 0$
defines an undirected weighted graph
$G(V,E)$, in which $V$ is the set of nodes
 and $E\subseteq V\times V$
is the set of edges
 describing the relation
 between pairs of nodes.
 }

 { The aim of this paper is to
investigate the properties
of the probabilistic locus satisfying
 (\ref{normalization}) (a graph)
specified by the arbitrary
symmetric, non-degenerated matrix
$w_{xy}\geq 0$
and to show that
it can be regarded
as the projective space
$P\mathbb{R}^{(N-1)}$
of homogeneous coordinates
obtained by projection
from the origin $\pi$.
Given a random walk
defined on the graph $G(V,E)$,
the projective geometry
can be reduced
 to the metric geometry,
which is realized by
 the introduction of
a Euclidean scalar product, a distance, and a norm
which can be directly related to
the first-access times of nodes
by random walkers.
}

 { The paper is organized as follows. In
(Sec.~\ref{sec:affinity}), we introduce the probabilistic analog
of affinity transformations that can be interpreted as random
walks defined on the set $V$ with the structure determined by the
matrix $w_{xy}\geq 0$. In (Sec.~\ref{sec:distances_and_angles}),
we consider the probabilistic projective geometry for the space
 projected from the distinguished probability distribution
$\pi$ induced by a simple stochastic process - a random walk -
instead a distinguished point as in the usual
projective geometry. In (Sec.~\ref{sec:metric}), we perform a
reduction from the probabilistic projective geometry to the
probabilistic metric geometry by introducing the Euclidean space
structure associated to the diffusion process defined on $V$ with
respect to  $w_{xy}\geq 0$. Then, in (Sec.~\ref{sec:FPT}), we
demonstrate that the Euclidean characteristics can be interpreted
in terms of the first-access properties of nodes by random
walkers. Using the metric structure introduced in
Sec.~\ref{sec:metric}, we introduce a probabilistic topological
space in Sec.~\ref{sec:topological}. We conclude in
Sec.~\ref{sec:examples} and give three visual representations of
the probabilistic locus of a chain, a polyhedron, and a finite
2-dimensional lattice. }

\section{ {Probabilistic analog of affinity transformations}}
\label{sec:affinity}
\noindent

  { In a geometric setting,
affine transformations \cite{Zwillinger} are precisely
the functions that map straight lines
 to straight lines, i.e.
preserves all linear combination
 in which the sum of the coefficients is 1.
 Their probabilistic analog
in the class of stochastic matrices,
$\sum_{y\in V}T_{xy}=1$,
 is given by
 the transition probability operators,
\begin{equation}
\label{transition}
T_{xy}\,=\,\left(1-\beta_x\right)\delta_{xy}+\frac{\,\,\beta_x \,w_{xy}\,\,}{\,\,\sum_{y\in V}w_{xy}\,\,}
\end{equation}
where $\beta_x\in\left[0,1\right]$
 and $w_{xy}\geq 0$ is the symmetric affinity
matrix describing
the set of paths
available from $x\in V$.
}

 {If points
of the discrete set $V$
are interpreted as nodes
of an undirected graph,
then
the transition operator
(\ref{transition}) describes a "lazy" random walk
on $G(V,E)$
specified by the "laziness" parameters $\beta_x$,
such that
 a random walker stays
in  $x\in V$
with probability $1-\beta_x$,
but moves
to another node
randomly chosen among the nearest neighbors
with
probability $\beta_x/\sum_{y}w_{yx}$.
In particular,
if  $\beta_x=1$ uniformly for all $x\in V$,
 the operator
(\ref{transition}) describes the usual
random walks
extensively studied in the classical surveys
\cite{Lovasz:1993},\cite{Aldous}.}

  { Random walks provide
us with an effective tool for the
detailed structural analysis
of connected undirected graphs exposing their
symmetries.
In \cite{Volchenkov},
we have shown that
 the transition operator
(\ref{transition})
gives the representation
of the group
of linear automorphisms
of the undirected graph $G(V,E)$,
in the class of
stochastic matrices. }

 {
Discrete translational symmetry
is obviously broken if
 the graph $G$
is either finite, or irregular.
If the graph is finite but regular,
then
the transition probability operator
(\ref{transition})
defines a symmetric Markov chain, in which
the probability of
moving to $y\in V$ given that a walker is
 at the node $x\in V$
is the same as the probability of
moving backward.
Finally, for a non-regular
graph $G$
this property is replaced
 by time--reversibility that
can be conveniently formulated
in terms of the
stationary random walks,
\begin{equation}
\label{time_reverse}
\pi_xT_{xy}\,=\,\pi_yT_{yx},
\end{equation}
 for every pair $x,y\in V$.
It is well known (see \cite{Lovasz:1993}) that the stationary
distribution of random walks $\pi$ defined on the undirected graph
$G$ is given by (\ref{stationary_pi}) and is the left eigenvector
of the transition operator (\ref{transition})
 belonging to
its primary Perron eigenvalue $\mu=1$.
Since
for any choice of $\beta_x\in [0,1]$
the matrix $T_{xy}\geq 0$
 is a real positive stochastic
matrix, it follows from  the
 Perron-Frobenius theorem (see \cite{Horn:1990})
 that its maximal eigenvalue $\mu=1$ is
 simple if the graph $G$ is connected.
Any other distribution $\sigma$ defined on $V$
such that  $\sigma_x\geq 0$ for any $x\in V$ and
 $\sum_{x\in V} \sigma _x=1$,
asymptotically tends to the stationary
distribution $\pi$
under the iterative
 actions of the transition operator  (\ref{transition}),
\begin{equation}
\label{limit}
\pi\,=\,\lim_{n\to\infty}\,\sigma\, T^{\,n}.
\end{equation}
 The self-adjoint operator associated to
$T$ is given by
\begin{equation}
\label{self_adj}
\widehat{T}\,=\,\frac 12\left( \pi^{1/2}T\pi^{-1/2}+\pi^{-1/2}T^{\top}\pi^{1/2}\right),
\end{equation}
where $T^\top$ is the adjoint operator, and $\pi$ is defined as
the diagonal matrix
$\mathrm{diag}\left(\pi_1,\ldots,\pi_N\right)$. While interesting
in the spectral calculations of random walks characteristics, the
symmetric matrix (\ref{self_adj}) is more convenient since its
eigenvalues are real and bounded in the interval  $\mu\in[-1, 1]$
and the eigenvectors define an orthonormal basis. }

\section{ {Probabilistic projective geometry}}
\label{sec:distances_and_angles}
\noindent

 { An affine coordinate system on $V$ is prescribed by an
affinely independent set of points for which the displacement
vectors $\mathbf{e}_y= {y}-{ x}$, $y\in V$, $x\ne y$,
  form a basis of $V$
with respect to the point $x\in V$. A displacement vector
$\mathbf{v}=\sum_{y\in V} v_y\mathbf{e}_y$ is identified with the
coordinate $(N-1)$-tuple $\left(v_1,\ldots,\{\,\,\}_x,\ldots,
v_N\right)$, in which the $x-$th component is missing. We can
associate all points $V$ with their relative displacement vectors.
}

 {This is
the distinguished
probability distribution $\pi$
induced by a simple stochastic process - a random walk -
is important for the defining
an elementary
coordinate system in the probabilistic
projective space.
Given a symmetric matrix $w_{xy}\geq 0$
and a vector $\beta_x\in [0,1]$,
we can define
 the transition probability
 by the kernel (\ref{transition})
on $V$ and
its self-adjoint counterpart (\ref{self_adj}).
The complete set of real
 eigenvectors $\Psi=\{\psi_1,\psi_2,\ldots \psi_N\}$
of the symmetric matrix (\ref{self_adj}),
\[
\left|\psi_i\right\rangle\widehat{T}\,=\,\mu_i\left|\psi_i\right\rangle,
\]
ordered in accordance to their eigenvalues,
 $\mu_1=1>\mu_2\geq\ldots\mu_N\geq -1$,
forms an orthonormal basis
in $\mathbb{R}^N$,
\begin{equation}
\label{orthonorm}
\left\langle\psi_x|\psi_y\right\rangle\,=\,\delta_{xy},
\end{equation}
 associated to linear automorphisms of
 the affinity matrix $w_{xy}$, \cite{Volchenkov}.
In (\ref{orthonorm}), we have used  Dirac's bra-ket notations especially
convenient for working with inner products and
rank-one
operators in Hilbert space.}

 { Given the random walk defined by the operator
(\ref{transition}), then the squared components of the
eigenvectors $\Psi$ have very clear probabilistic interpretations.
 The first eigenvector
    $\psi_1$ belonging to the largest eigenvalue $\mu_1=1$ satisfies
$\psi_{1,x}^2=\pi_x$
and describes the probability to find a random walker in $x\in V$.
The  norm
in the orthogonal complement of $\psi_1$,
$\sum_{i=2}^N\psi_{i,x}^2\,=\,1-\pi_x$,
is nothing else but the probability
that a random walker is not in  $x$.
}

 {Looking back it is easy to see that
the transition operator (\ref{self_adj}) defines
a projective transformation
on the set $V$ such that
all vectors in $\mathbb{R}^N(V)$
collinear to the stationary distribution
 $\pi>0$ (\ref{stationary_pi})
 are projected onto a common image point.
}

 {Geometric objects, such as points, lines, or planes,
can be given a representation as elements in projective
spaces based on homogeneous coordinates, \cite{Moebius}.
Any vector of the Euclidean space $\mathbb{R}^N$
can be expanded into $\mathbf{v}=\sum_{k=1}^N\left\langle\mathbf{v}|\psi_k\right\rangle\left\langle\psi_k\right|$,
as well as into the basis vectors
\begin{equation}
\label{homogen}
\psi'_s\equiv \left(1,\frac{\,\psi_{s,2}\,}{\,\psi_{s,1}\,},\ldots,
\frac{\,\psi_{s,N}\,}{\, \psi_{s,1}\,}
\right), \quad s\,=\,2,\ldots, N,
\end{equation}
 which span the
projective space $P\mathbb{R}_{\pi}^{(N-1)}$,
\[
\mathbf{v\pi}^{-1/2}\,=\,\sum_{k=2}^N\left\langle\mathbf{v}|\psi'_k\right\rangle\left\langle\psi'_k\right|,
\]
since we have always $\psi_{1,x}\equiv\sqrt{\pi_x}>0$ for any $x\in V$.
The set of all isolated vertices $p$ of the graph $G(V,E)$
for which $\pi_{p}=0$ play the role of the  plane at infinity,
 away from which
we can use the basis  $\Psi'$ as an ordinary Cartesian system. The
transition to the homogeneous coordinates (\ref{homogen})
transforms vectors of $\mathbb{R}^N$ into vectors on the
$(N-1)$-dimensional hyper-surface $\left\{
\psi_{1,x}=\sqrt{\,\pi_x\,}\right\}$, the orthogonal
complement to the
 vector of
stationary distribution $\pi$. }

 \section{ {Reduction to metric geometry}}
\label{sec:metric}
\noindent

\begin{figure}[ht]
 \noindent
\begin{center}
\epsfig{file=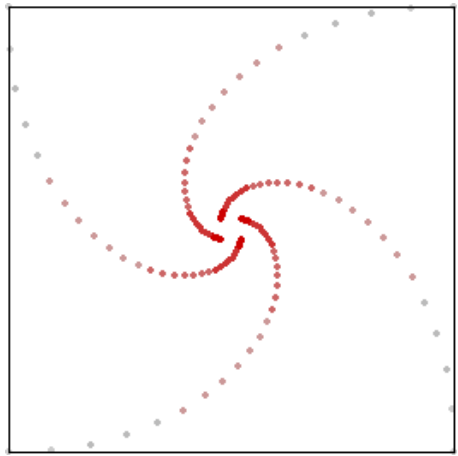, angle=0,width =6cm, height =6cm}
  \end{center}
\caption{\small  {Any vector $\xi\in
P\mathbb{R}_{\pi}^{(N-1)}$ asymptotically approaches  the origin
$\pi$ under the consecutive actions of the operator
$\widehat{T}$.}} \label{Fig1}
\end{figure}

 {
The key observation is that
in homogeneous coordinates
the operator
$\left.\widehat{T}^k\right|_{P\mathbb{R}_{\pi}^{(N-1)}}$ defined on
the
 $(N-1)$-dimensional hyper-surface
 $\left\{\psi_{1,x}=\sqrt{\,\pi_x\,}\right\}$
determines a contractive
discrete-time affine dynamical system.
The
origin
 is the only fixed point of the map $\left.\widehat{T}^k\right|_{P\mathbb{R}_{\pi}^{(N-1)}}$,
 \begin{equation}
\label{assymp}
\lim_{n\to\infty}\widehat{T}^{\,n}\xi\,=\, \left(1,0,\ldots 0\right),
\end{equation}
for any  $\xi\in P\mathbb{R}_{\pi}^{(N-1)}$
 and the solutions consist in the linear system
of  points
 $\widehat{T}^{\,n}\xi$
that hop in the phase space (see Fig.~\ref{Fig1}) along the curves
formed by collections of points that map into themselves under the
consecutive action of  $\widehat{T}$.}

 {
The problem of random walks (\ref{transition},\ref{self_adj})
defined on finite undirected graphs
can be related to a diffusion process which describes the
dynamics of a large number of random walkers.
The symmetric diffusion process correspondent
to the self-adjoint transition operator $\widehat{T}$
 describes the time evolution of the normalized expected number
of random walkers, ${\bf n}(t)\,\pi^{-1/2}\in V\times \mathbb{N}$,
\begin{equation}
\label{diffusion}
\dot{\bf n}\,=\,\widehat{ L}{\bf n},\quad \widehat{L}\,\equiv\, 1-\widehat{T}
\end{equation}
where $\widehat{L}$ is the normalized Laplace operator.
Eigenvalues of $\widehat{L}$ are
simply related to that of $\widehat{T}$,
$\lambda_k=1-\mu_k$, $k=1,\ldots, N$,
and the eigenvectors of both operators are identical.
The analysis of spectral properties of the operator
(\ref{diffusion}) is widely used in the spectral graph
theory, \cite{Chung:1997}.
}

 {
It is important to note that
the normalized Laplace operator (\ref{diffusion})
defined on $ P\mathbb{R}_{\pi}^{(N-1)}$
is invertible,
\begin{equation}
\label{inverted}
\begin{array}{lcl}
\widehat{L}^{-1}& =& \left(1-\widehat{T}\right)^{-1}\\
 &=& \sum_{n\geq 1} \widehat{T}^n.
\end{array}
\end{equation}
since $\left.\widehat{T}^k\right|_{P\mathbb{R}_{\pi}^{(N-1)}}$
is a contraction mapping for any $k\geq 1$.
The unique inverse operator,
\begin{equation}
\label{Green}
\widehat{L}^{-1}\,=\,
\sum_{s=2}^N\frac{\,\,\left|\psi'_s\right\rangle\left\langle\psi'_s\right|\,\,}{\,\,1-\mu_s\,\,},
\end{equation}
 is the Green function (or the Fredholm kernel)
describing long-range interactions between
eigenmodes of the diffusion process
induced by the
graph structure.
The convolution  with the Green's function gives solutions to
inhomogeneous Laplace equations.
}

 {In order to apply
metric geometry to the graph $G(V,E)$,
one needs to introduce the distances between points (nodes of the graph)
 and the angles between lines or vectors
that can be done by determining
the inner product
between
 any two vectors $\xi$ and $\zeta$ in $P\mathbb{R}_{\pi}^{(N-1)}$
as
\begin{equation}
\label{dot_product}
\left(\xi,\zeta\right)_{(w,\beta)} \,=\, \left(\xi,\widehat{L}^{-1}\zeta\right).
\end{equation}
The dot product (\ref{dot_product})
is a symmetric real valued scalar function that
allows us to define the (squared)
norm of a vector $\xi\in P\mathbb{R}_{\pi}^{(N-1)}$ with respect to
$(w,\beta)$ by
\begin{equation}
\label{sqaured_norm}
\left\|\, \xi\,\right\|^2_{(w,\beta)}\,=\,\left(\xi,\widehat{L}^{-1}\xi\right).
\end{equation}
The (non-obtuse) angle $\theta\in [0,180^{\mathrm{o}}]$ between two vectors
 is then given by
\begin{equation}
\label{angle}
\theta\,=\,\arccos\left(\frac{\,\,\left(\xi,\zeta\right)_{(w,\beta)} \,\,}{\,\, \left\|\, \xi\,\right\|_{(w,\beta)}\left\|\, \zeta\,\right\|_{(w,\beta)}\,\,}\right).
\end{equation}
The Euclidean distance between two vectors in
$P\mathbb{R}_{\pi}^{(N-1)}$ with respect to
$(w,\beta)$ is defined by
\begin{equation}
\label{spectral_dist}
\begin{array}{lcl}
\left\|\xi-\zeta\right\|^2_{(w,\beta)} &=& \left\|\, \xi\,\right\|^2_{(w,\beta)}+\left\|\, \zeta\,\right\|^2_{(w,\beta)}-2\left(\xi,\zeta\right)_{(w,\beta)}\\
 &=& \mathbb{P}_{\xi}\left(\xi-\zeta\right)+\mathbb{P}_{\zeta}\left(\xi-\zeta\right)
\end{array}
\end{equation}
where $\mathbb{P}_{\xi}\left(\xi-\zeta\right)\equiv\left\|\, \xi\,\right\|^2_{(w,\beta)}-\left(\xi,\zeta\right)_{(w,\beta)} $
and
$\mathbb{P}_{\zeta}\left(\xi-\zeta\right)\equiv\left\|\, \zeta\,\right\|^2_{(w,\beta)}-\left(\xi,\zeta\right)_{(w,\beta)} $
are the lengths of the projections of  $(\xi-\zeta)$ onto the unit vectors
in the directions of $\xi$ and $\zeta$ respectively.
 It is clear that
$\mathbb{P}_{\zeta}\left(\xi-\zeta\right)=\mathbb{P}_{\xi}\left(\xi-\zeta\right)=0$ if $\xi=\zeta.$
}

 { It is worth mentioning that
the spectral representations of the Euclidean structure
(\ref{dot_product}-\ref{spectral_dist}) defined for the graph nodes
can be easily derived by taking into account that
$\left\langle x\right.\left|\psi'_s\right\rangle=\psi'_{s,x}$.
}

\section{ {Euclidean structure and the first-access characteristic times of graph nodes}}
\label{sec:FPT}
\noindent

 {The structure of Euclidean space
  introduced in the previous
section can be related to a length structure $V\times V\to
\mathbb{R}_+$
 defined on a class
of all admissible paths $\mathcal{P}$
between pairs of nodes in $G$.
 It is clear that every path $P(x,y)\in \mathcal{P}$
is characterized by some probability
to be followed by
a random walker which  in particular depends upon
the weights $w_{pq}> 0$ of all edges $p\sim q$ that join it.
Therefore, the path length statistics
is a natural candidate for the length structure
on $G$.
 }

 { The theory of random walks on graphs
offers the concepts to quantify the
mutual accessibility of nodes in a graph.
The access time or hitting time $h_{xy}$ is the expected number of steps
before node $y$ is visited,  starting from node $x$, \cite{Lovasz:1993}.
Since the first step takes a walker to a neighbor
$v\sim x$ and then the walker has to reach
$y$ from there, the
hitting time should be additive,
\begin{equation}
\label{hitting}
h_{xy}\,=\, 1+ \frac {\,\,\sum_{x\sim v}h_{vy}\,\,}{\,\,\sum_{z\in V}w_{xz}\,\,}
\end{equation}
if $x\ne y$ and $h_{xx}=0.$
In general $h_{xy}$ is not a symmetric
matrix except
if the graph has a vertex-transitive automorphism group,
\cite{Lovasz:1993}. }

 {
The sum
\begin{equation}
\label{commute}
d(x,y)\,=\,h_{xy}+h_{yx}
\end{equation}
is called the commute time:
 this is the expected number of steps in a
random walk starting at $x$ before node $y$ is visited and then node $x$ is
reached again, \cite{Lovasz:1993}.
}

 {
The expected number of steps
 to first arrival
of
 a random walker
at
  $x\in V$
starting from another node randomly chosen with respect to the
probability $\pi$,
\begin{equation}
\label{FPT}
f_x\,=\,\sum_{y\in V}\pi_yh_{yx},
\end{equation}
is called the first-passage time to $x$.
 }

 { All three quantities
 are well-known in the
theory of random walks
on graphs and can be calculated
in the standard way, \cite{Lovasz:1993}.
The most important observation for us
is that
their spectral representations,
\begin{equation}
\label{spectral}
\begin{array}{lcl}
h_{xy}&=& \sum_{s=2}^N\,
\left(\psi'^2_{s,x}-\psi'_{s,x}\psi'_{s,y}\right)^2/\left(1-\mu_s\right),
\\
d(x,y)&=&
\sum_{s=2}^N\,
\left(\psi'_{s,x}-\psi'_{s,y}\right)^2/\left(1-\mu_s\right),
\\
f_x &=& \sum_{s=2}^N\,
\,\psi'^2_{s,x}\,/\left(1-\mu_s\right),
\end{array}
\end{equation}
coincide with those of the Euclidean quantities,
$\mathbb{P}_{x}(x-y)$, $\left\|x-y\right\|^2_{(w,\beta)} $, and
$\left\|x\right\|^2_{(w,\beta)} $ respectively. }

\section{ {Probabilistic  topological space}}
\label{sec:topological}
\noindent

 {
In the previous sections, we have shown that
given a symmetric affinity
function $w:V\times V\to \mathbb{R}_+$,
 we can always
define an Euclidean metric on $V$
based on the first-access properties
of the standard stochastic process, the random walks
defined on the set $V$ with respect to the matrix $w_{xy}\geq 0$.
}

 {
In particular, we can introduce
 this metric on any undirected graph $G(V,E)$
converting it in a metric space. The Euclidean distance
interpreted as the commute time induces the  metric
topology on $G(V,E)$. Namely, we define the open metric ball of
radius $r$ about any point $x\in V$ as the set
\begin{equation}
\label{ball}
B_r(x)\, = \,\left\{y \in V: d(x,y)< r\right\}.
\end{equation}
These open balls generate a topology on $V$, making it a topological space.
A set $U$ in the metric space is open if and only if for every point $x\in U$ there exists and
 $\varepsilon >0$ such that $B_\varepsilon(r)\subset U$, \cite{Burago}.
 Explicitly, a subset of $V$ is called open if it is a union of (finitely or infinitely many) open balls.
}

\section{ {Conclusion and examples}}
\label{sec:examples}
\noindent

 { Systems consisting of many individual units that are tied
by one or more specific types of interdependency are found
everywhere in the world. Networks are often very complex and
difficult to analyze. Being of rather large scale to be seen from
a single viewpoint, they can often be abstracted as graphs,
 the natural mathematical tool for facilitating the
analysis.}

\begin{figure}[ht]
\label{Fig2}
 \noindent
\begin{center}
\begin{tabular}{llrr}
 1. &\epsfig{file=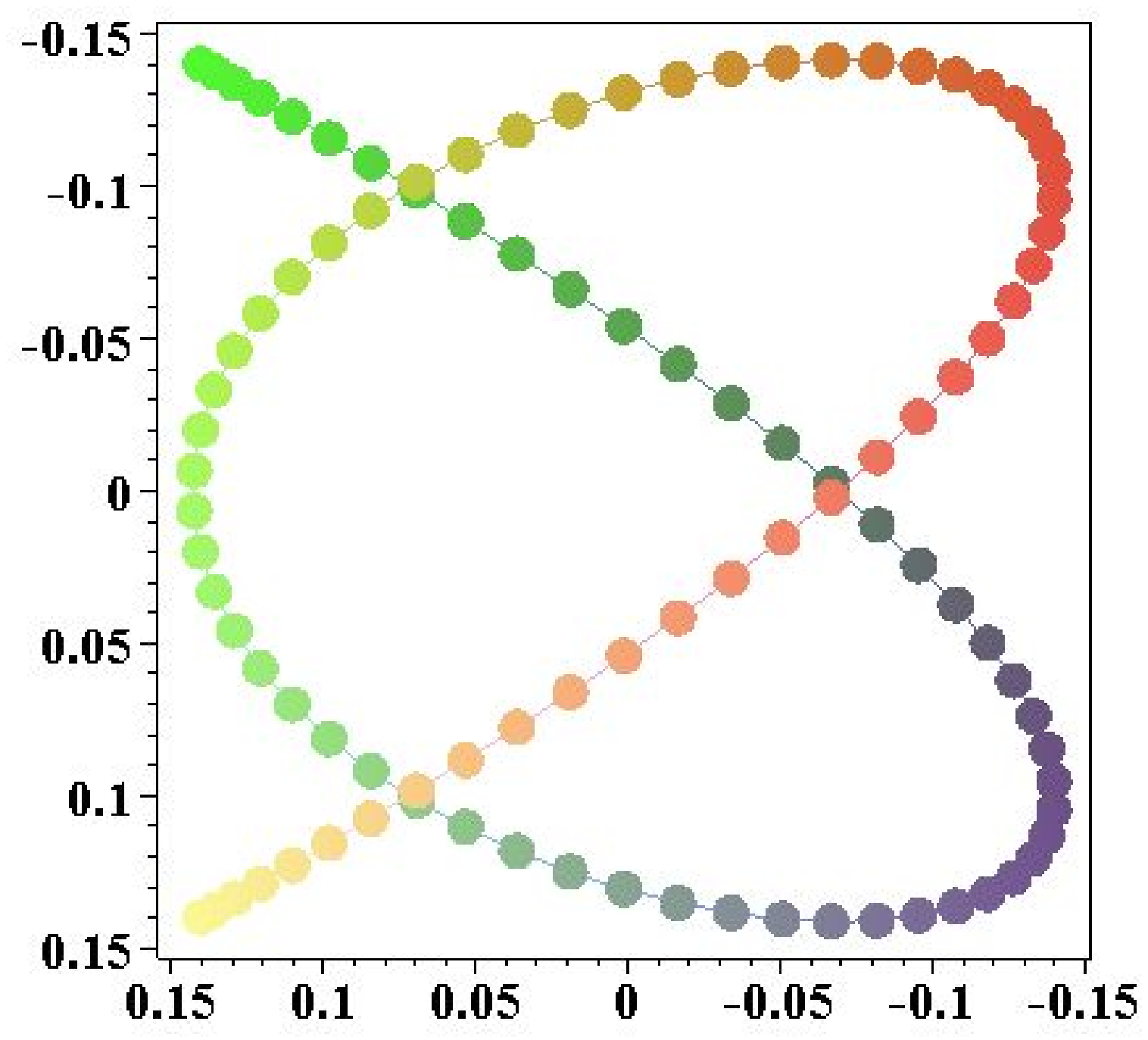, width=6.0cm, height =5.5cm}&2.&
 \epsfig{file=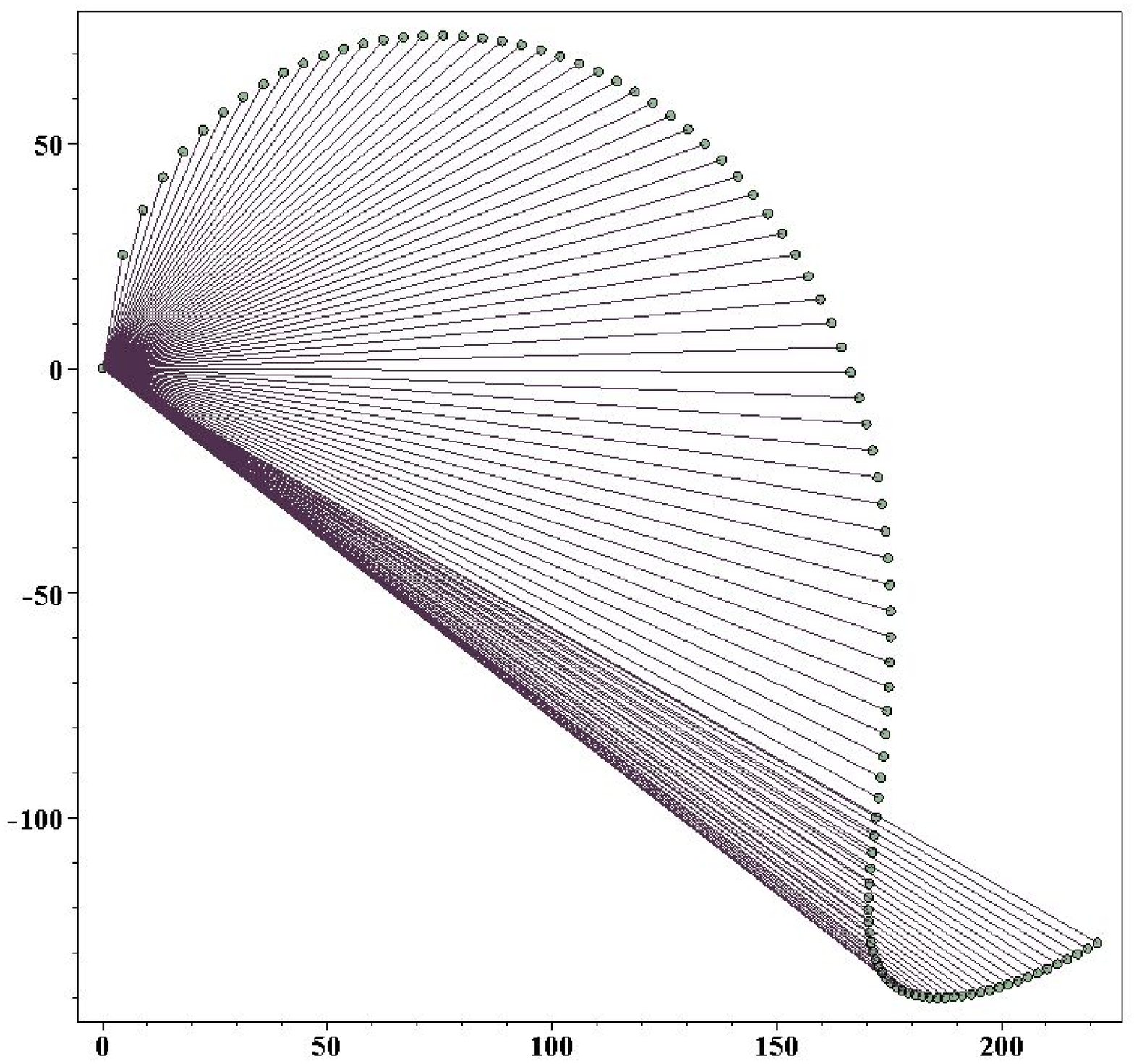, width=6.0cm, height =5.3cm} \\
 3. &\epsfig{file=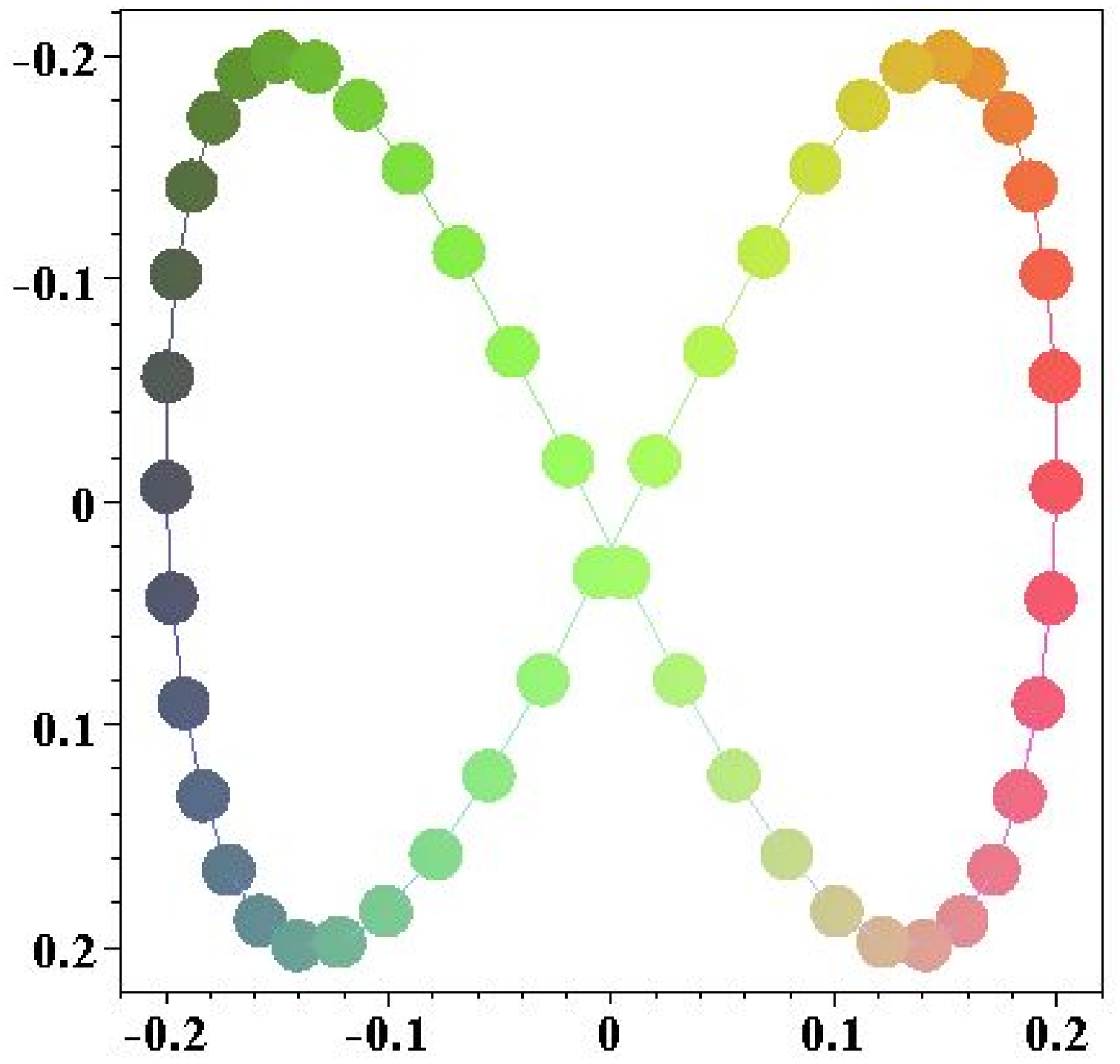,width=6.0cm, height =5.5cm}&4.&
 \epsfig{file=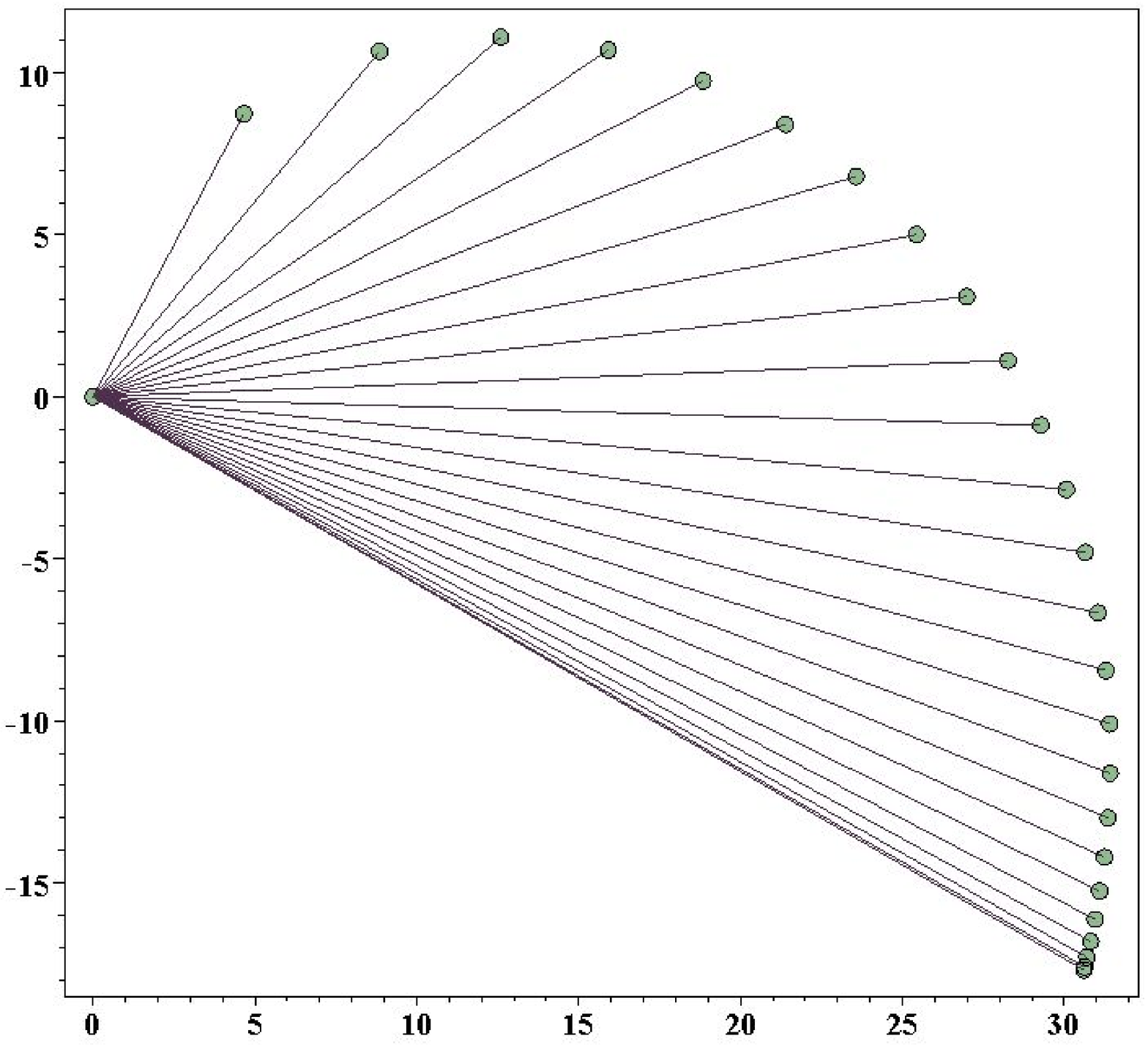, width=6.0cm, height =5.5cm}
\end{tabular}
\end{center}
\caption{ {Probabilistic images of  a chain ($N=100$) and a polyhedron ($N=50$).
}}
\end{figure}

 {Euclidean space has a decisive role in visual and
propriomotor percepts, and in hearing thus determining our spatial
perception. In addition, many of our feelings, of anger, fear and
so on, have important links with parts of the body and hence
indirectly with Euclidean space. In general, networks do not
possess the structure of Euclidean space. Thus, a mental
representation of any network emerges as a result of a long
learning process jointly with the planning of movements in that
and is always challenging  to make the proper decisions on how to
sustain the system and to cope with the new demand. }

\begin{figure}[ht]
\label{Fig3}
 \noindent
\begin{center}
\begin{tabular}{llrr}
 1. &\epsfig{file=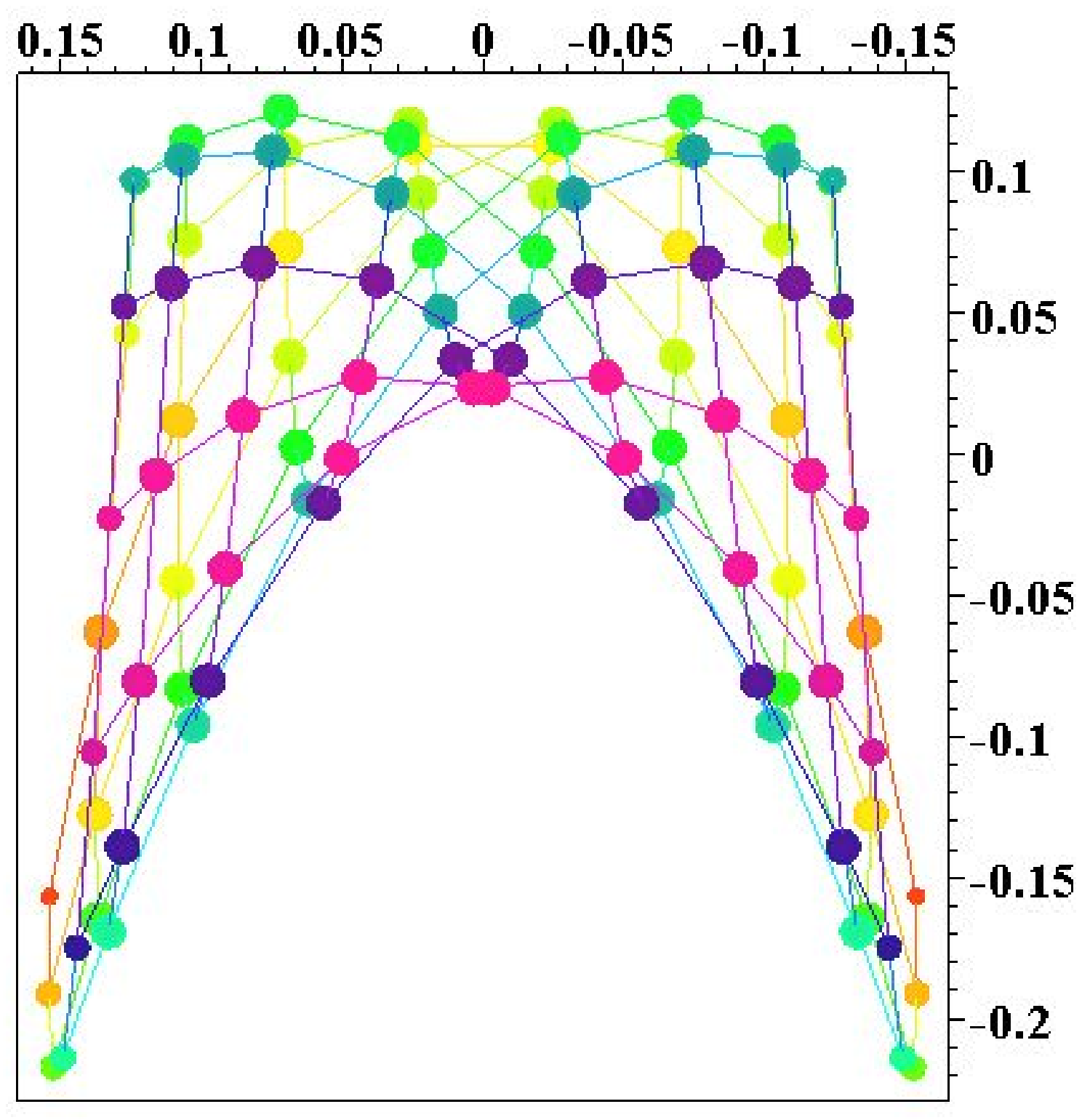, width=6.0cm, height =6.0cm}&2.&
 \epsfig{file=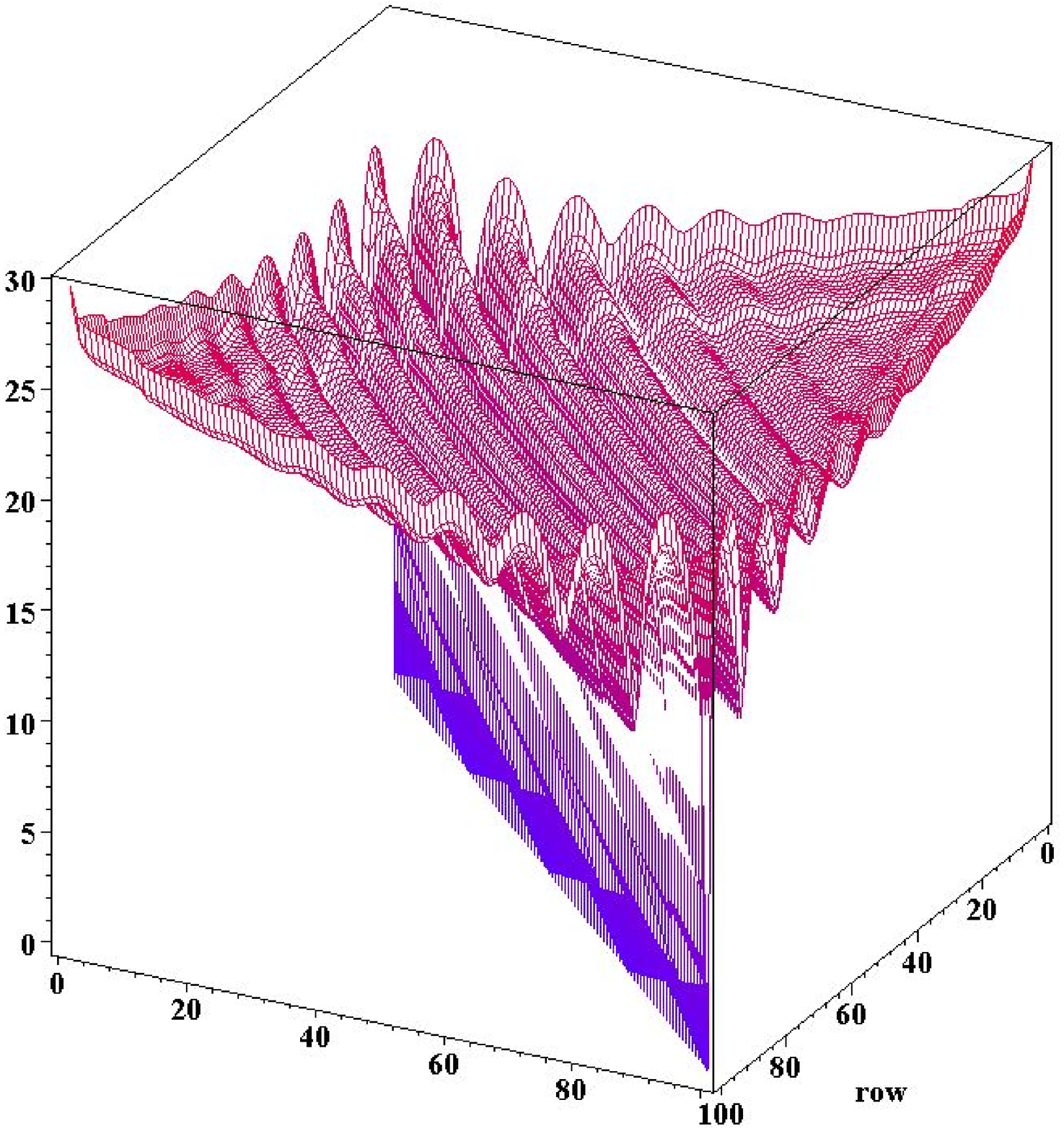, width=6.0cm, height =7.0cm}
\end{tabular}
\end{center}
\caption{1. Probabilistic image of a 2D-lattice ($N=10^2$).
2. The matrix plot of the probabilistic Euclidean distances (the commute times) between the
nodes of the lattice $\mathcal{L}_2$.}
\end{figure}

 {
In our paper, we have demonstrated that
networks do possess the structure of Euclidean space,
 but in the probabilistic sense.
}

 {
In order to illustrate the approach, we
have shown the probabilistic images of
a chain (1D-lattice of $N=100$ nodes) (see Fig.~2(1,2)), a polyhedron
(a cycle of $N=50$ nodes) (see Fig.~2(3,4)), and a 2D-lattice $\mathcal{L}_2$ containing
 $10^2$ nodes (see Fig.~3). We suppose that all
weights are equal $w_{xy}=1$, so that the respective affinity
matrices are just the adjacency matrices of the graphs. }

 { Random walks defined on the above networks embed them
into the $(N-1)$-dimensional locus of Euclidean space, in which
all nodes acquire  certain norms quantified by the
first-passage times to them from  randomly chosen nodes.
Indeed, the structure of $(N-1)$-dimensional vector spaces induced
by random walks cannot be represented visually.}

 {In order to obtain a 3D visual representation
of these graphs, we  have calculated their three major
eigenvectors $\{\psi_2,\psi_3,\psi_4\},$  belonging to the largest
 eigenvalues
$\mu_k<1$ of the symmetric transition operators (\ref{self_adj}).
  The $({\bf e}_1,{\bf e}_2,{\bf e}_3)$-coordinates
of the vertex $x\in V$ of the graph
in 3D space have been taken equal to the relevant $x^\mathrm{th}$-components of three
eigenvectors $\{\psi_2,\psi_3,\psi_4\}$.
  The radiuses of balls
representing nodes in Figs.~2,3 have been taken proportional to
the degrees of nodes. In Figs.~2.(1,3) and in Fig.~3(1), we have
presented the the 3D images of the the chain, polyhedron , and the
lattice $\mathcal{L}_2$. The connections between nodes represent
the actual connections between them in the real space. }

 {If we choose
one node of a graph
as a point of reference, we
can draw the
2-dimensional projection of the $(N-1)$-dimensional
locus
by arranging other nodes at
the distances calculated
accordingly to (\ref{commute}) and under
the angles (\ref{angle}) they
are with respect to the
chosen reference node.
The examples are given in Fig.2(2,4).
}

 { In particular, in Fig.2(2), we have presented the
2-dimensional projection of the 99-dimensional Euclidean locus for
a chain with respect to the marginal left node. The probabilistic
Euclidean distance measured by the commute time (\ref{commute})
from the left end node
 is increasing node by node
approximately as $\propto\sqrt{N}$ from  $18.166$ random steps for
the nearest neighbor node
 to $180.748$ random steps for the node at the opposite end of the chain
($x=100$). In Fig.2(4), the similar diagram is represented for the
polyhedron. It is worth to mention that the symmetry of the
polyhedron can also be seen in the image of its $49-$dimensional
probabilistic locus. Chosen a vertex of the polyhedron as the
reference node, the commute time with its nearest neighbors equals
$9.899$ random steps, while it takes in average $35.355$ random
steps in order to commute with the vertex  on the circumcircle
diametrically opposite to the origin.
 }

 {
In Fig.3(2), we have shown the matrix plot of the
 probabilistic Euclidean distances (the commute times) between all
nodes of the lattice $\mathcal{L}_2$. The distances on the diagonal $d(x,x)=0$,
and vary harmonically from 15 to 30 random steps for different pairs of nodes.
}

\section{ {Acknowledgment}}
\label{Acknowledgment}
\noindent

 {
This work has been supported by the Volkswagen Foundation (Germany)
in the framework of the project "Network formation rules, random
set graphs and generalized epidemic processes" (Contract no Az.:
I/82 418).}

\end{document}